\documentclass[prl,aps,twocolumn,superscriptaddress,floatfix,nofootinbib,showpacs,longbibliography]{revtex4-1}

\usepackage[utf8]{inputenc}  
\usepackage[T1]{fontenc}     
\usepackage[british]{babel}  
\usepackage[sc,osf]{mathpazo}
\usepackage[scaled=0.86]{berasans}  
\usepackage[colorlinks=true, citecolor=blue, urlcolor=blue]{hyperref}  
\usepackage{scrextend}
\usepackage{graphicx} 
\usepackage[babel]{microtype}  
\usepackage{amsmath,amssymb,amsthm,bm,amsfonts,mathrsfs,bbm} 

\usepackage{xspace}  
\usepackage{pgf,tikz}
\usepackage{xcolor,colortbl}
\usepackage{array}
\usepackage{bigstrut}
\usepackage{braket}
\usepackage{color}
\usepackage{natbib}
\usepackage{slashbox}

\newcommand{\be}{\begin{equation}}
\newcommand{\ee}{\end{equation}}
\newcommand{\ba}{\begin{eqnarray}}
\newcommand{\ea}{\end{eqnarray}}

\begin{document}

\title{Exclusivity principle and unphysicality of Garg-Mermin correlation} 
\author{S. Aravinda}
\email{aravinda@imsc.res.in}    
\affiliation{Optics and Quantum Information Group, The Institute of Mathematical Sciences, HBNI, CIT Campus,
Taramani,  Chennai  600113, India}
  
\author{Amit Mukherjee}
\email{amitisiphys@gmail.com}    
\affiliation{Physics and Applied Mathematics Unit, Indian Statistical Institute, 203 B. T. Road, Kolkata 700108, India}

\author{Manik Banik}
\email{manik11ju@gmail.com} 
\affiliation{Optics and Quantum Information Group, The Institute of Mathematical Sciences, HBNI, CIT Campus,
Taramani,  Chennai  600113, India}

\begin{abstract}
The question concerning 
the  physical  realizability of  a probability
distribution is of quite importance in Quantum foundations.
Specker first pointed out  that this question cannot be
answered  from Kolmogorov's  axioms alone.  Lately,  this
observation of  Specker has  motivated simple  principles (exclusivity
principle/  local orthogonality  principle) that  can explain  quantum
limit regarding the possible  sets  of  experimental   probabilities  in
various nonlocality and  contextuality   experiments.  We   study  Specker's
observation in the simplest scenario  involving three inputs each with
two  outputs. Then  using  only linear  constraints  imposed on  joint
probabilities by this principle, we reveal unphysical nature of Garg-Mermin (GM) correlation. 
Interestingly, GM correlation was proposed to falsify the following suggestion by Fine:  if the inequalities of  Clauser and Horne (CH) holds,  then there
exists a deterministic local hidden-variable model for a  spin-1/2 correlation experiment
of the Einstein-Podolsky-Rosen type, even when more than two observables are involved on each side. Our result establishes that, unlike in the 
CH scenario, the local
orthogonality  principle at  single copy  level is  not equivalent to  the
no-signaling condition in the GM scenario.
\end{abstract}



\maketitle

\emph{Introduction}  -- The  origin of  probability theory,  like many
other branches of mathematics, lies in physical observation associated
with  the  occurrence  of  different possible  events/outcomes  of  an
experiment,  viz., coin  tossing, rolling  a  dice etc.  With time  it
becomes an increasingly large area  of study in mathematics which also
finds  useful applications  in different  areas of  physics-- starting
from the  study of Brownian  motion to statistical physics  to quantum
mechanics.

In  this article we study  a very naive question  that goes
back to the origin  of the relation between mathematics and a physical theory, in particular, 
\emph{physical realizability}
of  a  mathematically  well  defined  probability  distribution. More
explicitly we focus  on the question whether  an arbitrary probability
distribution satisfying Kolmogorov's axioms  can always be realized in
some physical experiment.  Specker first noticed that  the question of
physical  realizability  of  probability   distributions  can  not  be
answered from  Kolmogorov's axioms alone \cite{Spek60} (see \cite{Seevinck11} for English translation).  He actually pointed
out that if each pair of  a set of experimental outcomes are exclusive
alternatives  in  some  measurement,   then  their  probabilities  are
consistent with the  existence of a further measurement  in which they
are all  exclusive. This observation was  used by Wright to  show that
simple  sets of  events allow  probabilities such  that their  sum can
exceed the maximum  classical value \cite{Wri78}, and this  concept is also
strongly related to \emph{orthomodularity} and \emph{orthocoherence} \cite{FR81}.

In the recent past, this  particular observation of Specker has gained
renewed  interest  and  different  researchers  have  studied  closely
related  version  of  it   under  different  names:-  exclusivity  (E)
principle \cite{Cab13b}, local orthogonality (LO) principle (when applied to
nonlocality)   \cite{FSA+13},   consistent    exclusivity   (CE)   principle
\cite{Hen12}.  These principles  have proved  to be  useful towards  a more
unified  understanding of  quantum contextuality \cite{K67,KCB+08}  and nonlocality \cite{Bel64,BellRMP66,MerminRMP93,BCP+14},  in
particular, nonlocal scenario involving more  than two parties \cite{Cab13b,FSA+13,Hen12,Subh13,Bin13,Cab14}. 
Different other information theoretic/physical principles like information causality (IC)
\cite{Paw+09}, macroscopic  locality (ML) \cite{Nav10} along with other methods \cite{Oppenheim10,Banik13,Gazi2013,Das13} have  also been introduced
to point out unphysical correlations in the bipartite scenario. The LO
principle, likewise  IC and ML,  serves the same purpose  in bipartite
scenario, but its domain of  applicability is not limited to bipartite
scenario alone  as in the  case of IC and  ML \cite{FSA+13}. For  the single
party scenario  E-principle  has also  been successfully  used to
explain limited  noncontextual behavior of quantum  theory \cite{Cab13b}. On
the  other hand,  in  a  recent work  \cite{Hen15}  Henson has  established
interesting  connection  between  this  principle  and  a  generalized
version  of  Sorkin's  idea  of  `lack  of  (irreducible)  third-order
interference' \cite{Sor94}.

While studying the  nonlocality scenario, the authors  in \cite{FSA+13} have
proved (originally shown in \cite{Winter10,Seve14}) that the  constraints imposed by LO principle  on the bipartite
correlation  is equivalent  to the  no-signaling (NS)  constraints and
hence  linear on  joint  probabilities. However,  LO principle,  while
applied to several  copies of a device imply  non-linear constraint on
the joint probability  distribution and turns out to  be very powerful
for  ruling out  non-quantum correlations, viz, the canonical Popescu-Rohrlich (PR) correlation \cite{PR94}. Interestingly,  in
this article we  show that the equivalence between single  copy LO and
NS as established  in \cite{FSA+13,Winter10,Seve14} is restrictive in a  sense. We show that
there exists  situation where the  constraint imposed by LO  at single
copy  level can  rule out  non-quantum correlations  even though  they
satisfy  the NS  conditions.  To  show this  we  consider  a class  of
bipartite correlations introduced by Garg and Mermin (GM) \cite{GM82}. The
GM correlation has been proposed as a counterexample for the suggestion
by Fine  \cite{Fin82,FineRep} that if the  inequalities of Clauser and  Horne hold \cite{CH74} for each set of four observables, two from each side
then there exists  a hidden-variable model for  a spin-1/2 correlation
experiment of  the Einstein-Podolsky-Rosen \cite{EPR35} type even when more than two observables are considered on each side. Though  Horodecki and Horodecki
in Ref. \cite{HH96}  pointed out that GM correlations cannot be 
realizable by bi-partite two-qubit   quantum mechanical state,
 they did
not  exclude  a  more   sophisticated  experiments  for  the  physical
realization of GM correlations. We prove the strongest no-go result in this
direction by  showing   that   GM  correlations cannot be realizable in any physical theory. 
The unphysicality of GM correlations is established using
the constraints imposed by E-principle at single copy level only.
\begin{figure}[t!]
	\centering
	\includegraphics[height=4cm,width=7cm]{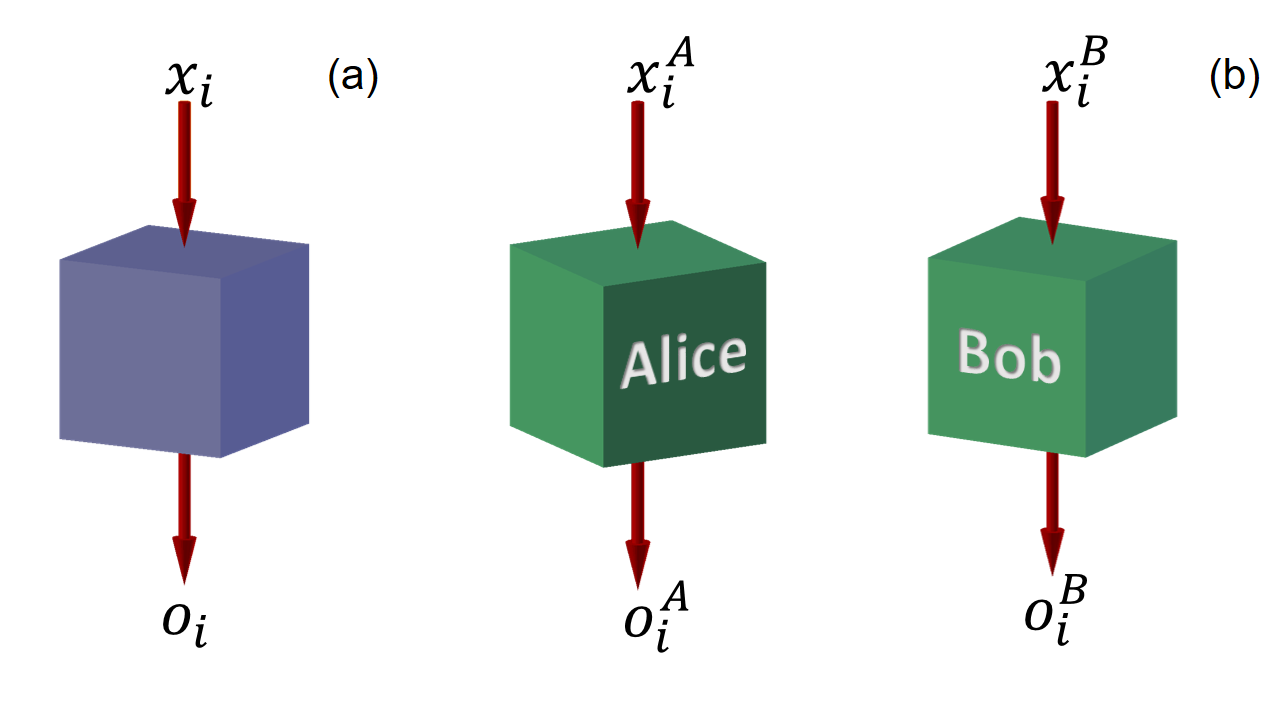}
	\caption{(a)  Single-party black  box scenario.  $x_i$ denotes
		input  and  correspondingly  $o_i$ the  output.  E-principle
		concerns   the  scenario   whenever  a   joint  distribution
		$\{P(o_1...o_k|x_1...x_k)\}$  exists  for  a set  of  inputs
		$\{x_i\}_{i=1}^k$.   In  general   the   output  $o_i$   are
		multivalued,  but  here  we  consider  $o_i\in\{0,1\}$.  (b)
		Two-party  black box  scenario. Joint  distributions of  the
		type  $\{P(o^A_io^B_j|x^A_ix^B_j)\}$  always exist  for  all
		$i,j$.}\label{fig1}
\end{figure}


\emph{The  set up}  -- We  consider  the `black  box' framework  which
contemplates  the probability  distributions  on the  outputs/outcomes
given some  inputs/measurements and also considers  correlations among
outcomes  of measurements  in  multipartite  scenarios. This  approach
makes  no  assumptions on  the  internal  working of  the  measurement
devices (hence the name `black  box' framework) and also  it does
not  consider   the  post  measurement   state  of  the   system  [See
  Fig.\ref{fig1}].  By  $e=(o_1o_2...o_n|x_1x_2...x_n)$ we  denote  an
event that  the outcomes  $o_1,o_2,...,o_n$ are  respectively obtained
when  the compatible  tests $x_1,  x_2, ...  ,x_3$ are  performed. For
multipartite scenario  superscripts will  be used to  denote different
parties,                                                         e.g.,
$(o^A_1o^A_2...o^B_1o^B_2...|x^A_1x^A_2...x^B_1x^B_2...)$,       where
$x^A_i$'s ($x^B_i$'s) denotes inputs of Alice (Bob) with corresponding
outputs  $o^A_i$  ($o^B_i$).  Whenever   there  is  no  confusion  the
superscript  will  be  avoided  in  general.  Two  events  are  called
exclusive   if    they   cannot   be   simultaneously    true,   i.e.,
$e=(o_1o_2...o_n|x_1x_2...x_n)$                                    and
$e'=(o'_1o'_2...o'_n|x'_1x'_2...x'_n)$ are  exclusive if for  some $i$
(at  least   one)  $o_i\neq  o'_i$  whereas   $x_i=x'_i$. Exclusivity
principle  places a constraint   that  the  sum  of  the  probabilities  of  pairwise
exclusive events $\{e_j\}$ cannot exceed unity, i.e., $\sum_jP(e_j)\le
1$.

\emph{Three-input two-output  scenario} -- Consider the  scenario with
three    inputs   $\{x_1,x_2,x_3\}$    each   with    binary   outputs
$o_1,o_2,o_3\in\{0,1\}$.  Also consider  that all  the three  bi-joint
probability distributions  $\{P(o_io_{j}|x_ix_{j})\}_{i<j}$ exist. The
tri-joint probability distribution $\{P(o_1o_2o_3|x_1x_2x_3)\}$ may
or  may  not  exist.  We also  consider  the  probability  consistency
condition, i.e.,
\begin{equation}
P(o_1|x_1):=\sum_{o_2}P(o_1o_2|x_1x_2)=\sum_{o_3}P(o_1o_3|x_1x_3),
\end{equation}  
and  similar   conditions  for  the  other   single  marginals.  These
conditions are also  known as no-disturbance (ND)  principle \cite{Ramanathan12}. 
General three bi-joint probability  distribution box satisfying the ND
principle can be expressed as in the Table-\ref{tab:bimar}.

\begin{table}[t]
	\scalebox{1.02}{
	\begin{tabular}{|c||c|c|c|c|}
		\hline
	\backslashbox{input}{output}& 00 & 01 & 10 & 11 \\
		\hline
		$x_1x_2$ & $c_{12}$ & $m_1-c_{12}$ & $m_2-c_{12}$ & $1-m_1-m_2+c_{12}$ \\
		\hline
		$x_2x_3$ & $c_{23}$ & $m_2-c_{23}$ & $m_3-c_{23}$ & $1-m_2-m_3+c_{23}$ \\
		\hline
		$x_1x_3$ & $c_{13}$ & $m_1-c_{13}$ & $m_3-c_{13}$ & $1-m_1-m_3+c_{13}$ \\
		\hline
	\end{tabular}}
	\caption{Three-input two-output bi-joint ND probability distributions, where $m_i:=P(o_i=0|x_i)$ and $c_{ij}:=P(o_i=o_j=0|x_ix_j)$. Positivity conditions imply for all the element in the table.}	\label{tab:bimar}
\end{table}

We are interested in the question  whether such a probability box will
be  realized in  some physical  scenario. One  trivial answer  to this
question  is that if a tri-joint probability  distribution
$\{P(o_1o_2o_3|x_1x_2x_3)\}$ 
exists which  reproduces all the three bi-joint distributions in the above table,
then  the  box  is  realizable  even with a  physical  experiment  in 
classical world. 

However, existence  of three-joint probability is not
a necessary condition for physical realizability of such a probability
box. For example, consider unsharp measurements on a two-level quantum
system.  With suitable  value of  unsharpness parameter  one can  find
three  observables  that  are  pairwise  jointly  measurable  but  not
triple-wise jointly  measurable. In  such a scenario  one can  obtain a
probability box with no tri-joint  probability but having a physical
realization \cite{Ravi14}.

E-principle provides non-trivial constrains  that need to be satisfied
by the probability box of  Table.\ref{tab:bimar} for having a physical
realization. In the single copy level these constrains are,
\begin{equation}\label{ex}
P(o_1o_2|x_1x_2)+P(\bar{o}_2o_3|x_2x_3)+P(\bar{o}_1\bar{o}_3|x_1x_3)\le1,
\end{equation}
where $\bar{o}_i$ denotes complement of  the output $o_i$. Varying the
outputs  $\bar{o}_1$,  $\bar{o}_2$,  and   $\bar{o}_3$  we  get  eight
different    inequalities.      E-principle    while     applies    to
$k\in\mathbb{Z}_+$ copy  level imposes further non  trivial constraint
(in general  nonlinear on  probabilities) for  $k>1$ \cite{FSA+13}.  In this
article we mostly consider the  single copy scenario. Whenever a given
probability box violates any  of the above inequalities-(\ref{ex}), it
become illegitimate  to be  a physical probability  distribution. Note
that whenever a probability  box of  Table-\ref{tab:bimar} is obtained from a three-joint distribution as  marginal then it satisfies all the
constraints imposed by the E-principle \cite{app}.

The set of three-input two-output  distributions satisfying ND forms a
six dimensional polytope $\mathcal{ND}$.  Positivity conditions of the
probabilities  configure   the  facets   of  $\mathcal{ND}$.   Set  of
probability distributions  satisfying E-principle  at k-level  forms the set $\mathcal{E}^k$ (not convex in general), with the following set inclusion relation:
$...\mathcal{E}^{k+1}\subseteq\mathcal{E}^{k}\subseteq\mathcal{E}^{k-1}\subseteq...\mathcal{E}^{1}\subseteq\mathcal{ND}$ \cite{Acin2015}. There
exist   eight   deterministic   boxes   satisfying   all   the   above
inequalities-(\ref{ex}),  and   four  non  deterministic   boxes  each
violating  two of  the  exclusivity inequalities-(\ref{ex})  maximally
\cite{app}.  These  twelve  boxes   (eight  deterministic  and  four  non
deterministic)  are   the  extremal   points  of   the  $\mathcal{ND}$
polytope. Note that while the algebraic value of the left hand side of
the  expressions in  Eq.(\ref{ex})  can  takes value  up  to $3$,  the
maximal value satisfying ND is limited to $3/2$. This is unlike to the
bipartite scenario  of Clauser-Horne-Shimony-Holt (CHSH) inequalities \cite{CHSH}, where  the algebraic optimal
value  of the  CHSH expression  can be  achieved by  a NS  correlation
\cite{PR94}. Though  $\mathcal{E}^{k}$'s are in general  convex sets, only
$\mathcal{E}^{1}$    is    a    polytope   lying    strictly    within
$\mathcal{ND}$. The non-trivial facets  of $\mathcal{E}^{1}$ are given
by the inequalities-(\ref{ex}).

From the above  discussion it is evident that  the constraints imposed
by  E-principle   at  single   copy  ($k=1$)   level  are   linear  on
probabilities, while they are in general non linear for higher $k$ and
hence   $\mathcal{E}^{k>1}$'s    are   not   polytopes    but   convex
sets. Moreover, for single party  scenario $\mathcal{E}^1$ is a proper
subset of $\mathcal{ND}$. However,  for bipartite scenario the authors
in Ref.\cite{FSA+13} have shown that E-principle (or LO principle) at single
copy level  and the NS principle  define same set of  correlations. In
what follows  we show that this  result is restrictive in  a sense. We
find example of bipartite correlations  that satisfy NS conditions but
violate the  linear constrains  imposed by  E-principle at  single copy
level. To prove this claim we use a correlation that was introduced by
Garg and Mermin  in a comparatively old paper  \cite{GM82}. Before proving
our claim, let us first digress on GM correlation a bit.

\emph{Garg-Mermin correlation} -- Arthur Fine in one of his well known
paper established that satisfaction of Bell/Clauser-Horne inequalities
is necessary  and sufficient conditions  for the existence of  a joint
probability distribution for the four observables (two on Alice's side
and two on  Bob's side) involved in the inequalities  \cite{Fin82}. He also
suggested that the Bell/Clauser-Horne inequalities would be sufficient
for more general scenario involving  more than two observables on each
side  \cite{Fin82}.  Garg and  Mermin  proved Fine's claim invalid by
providing  an  explicit  set  of pair  distributions  involving  three
observables on each side. The      GM     correlation
$\mathcal{GM}(c)\equiv\{P(o^A_io^B_j|x^A_ix^B_j)~|~o^A_i,o^B_j\in\{0,1\};~i,j\in\{1,2,3\}\}$
is given in the Table-\ref{tab:gm}.
\begin{table}[t]
\scalebox{1.2}{	
	\begin{tabular}{|c||c|c|c|c|}
		\hline
	\backslashbox{in}{out}& $0^A0^B$ & $0^A1^B$ & $1^A0^B$ & $1^A1^B$ \\
		\hline
				& & & &\\[-.3cm]			
		$x^A_1x^B_1/x^A_2x^B_2$ & $(1-c)/2$ & 0 & 0 & $(1+c)/2$ \\
		\hline
		& & & &\\[-.3cm]
		Else & $(1-3c)/6$ & $1/3$ & $1/3$ & $(1+3c)/6$\\
				\hline
	\end{tabular}}
	\caption{GM correlations $\mathcal{GM}(c)$, where $0<c\le 1/3$.}	\label{tab:gm}
\end{table}
The marginal probability distributions for  Alice and Bob are same and
read  as   $P(0|x^{\mu}_i)=(1-c)/2$  for  $\mu\in\{A,B\}$   and  $i\in
{1,2,3}$.  Clearly,  $\mathcal{GM}(c)$  satisfies  the  following  two
features:       (i)       Joint       probability       distribution
$\{P(o^{\mu}_io^{\mu}_j|x^{\mu}_ix^{\mu}_j)\}$  exists   for  all  the
three  pairs $ij\in\{12,23,13\}$ and $\mu\in\{A,B\}$.  
(ii)
The           three-joint           probability           distribution
$\{P(o^{\mu}_1o^{\mu}_2o^{\mu}_3|x^{\mu}_1x^{\mu}_2x^{\mu}_3)\}$  does
not exist,  otherwise $\mathcal{GM}(c)$ will  turns out to be  a local
correlation \cite{GMcommentFine}.

According to  the property  (i), the  most general  form of  the three
bi-joint  probability distribution  on Alice's/Bob's  side reproducing
single  marginals compatible  with $\mathcal{GM}(c)$  is given  in the
Table-\ref{tab:bimar1}.
\begin{table}[b]
	\scalebox{1.15}{
	\begin{tabular}{|c||c|c|c|c|}
		\hline
		\backslashbox{in}{out}& 00 & 01 & 10 & 11 \\
		\hline
		& & & &\\[-.3cm]
		$x^{\mu}_1x^{\mu}_2$ & $\alpha$ & $(1-c)/2 - \alpha$ & $(1-c)/2 - \alpha$ & $c+\alpha$ \\
		\hline
		& & & &\\[-.3cm]
		$x^{\mu}_2x^{\mu}_3$ & $\beta$ & $(1-c)/2 - \beta$ & $(1-c)/2 - \beta$ & $c+\beta$ \\
		\hline
		& & & &\\[-.3cm]
		$x^{\mu}_1x^{\mu}_3$ & $\gamma$ & $(1-c)/2 - \gamma$ & $(1-c)/2 - \gamma$ & $c+\gamma$ \\
		\hline
	\end{tabular}}
	\caption{Three bi-joint probability distribution, where $\mu=A,B$. Positivity conditions imply $0\le\alpha,\beta,\gamma\le (1-c)/2$.} \label{tab:bimar1}
\end{table}
Given  three  single  marginal  distributions  $\{P(o_i|x_i)\}$,  with
$i=1,2,3$,  one  can  always   construct  a  three-joint  distribution
$\{P(o^{\mu}_1o^{\mu}_2o^{\mu}_3|x^{\mu}_1x^{\mu}_2x^{\mu}_3)\}$
reproducing  the  marginals.  Actually the  solution  for  three-joint
distribution  is not  unique and  one trivial  example is  the product
distribution.   However,   given   the  three   bi-joint   probability
distributions as in Table-\ref{tab:bimar1}  existence of a three-joint
distribution   reproducing  the   bi-joints   as   marginals  is   not
guaranteed.  A general  problem of  this kind  is addressed  by Farkas's
lemma \cite{Garg84}. For  our case a straightforward  calculation shows that
probability box  as in Table-\ref{tab:bimar1}  can be obtained  from a
three-joint                  probability                  distribution
$\{P(o^{\mu}_1o^{\mu}_2o^{\mu}_3|x^{\mu}_1x^{\mu}_2x^{\mu}_3)\}$    as
marginals \emph{iff} the following conditions are satisfied:
\begin{eqnarray}\label{eq-3}
\alpha + \beta + \gamma & \geq \frac{1-3c}{2}; 
~~~~~~\beta + \gamma  \leq \frac{1-3c}{2}; \nonumber\\
\alpha + \beta  & \leq \frac{1-3c}{2}; 
~~~~~~\alpha + \gamma  \leq \frac{1-3c}{2}.
\end{eqnarray}
Similar conditions have also been obtained by A. Fine in Ref.\cite{AFine}. According to the property (ii) of the GM correlation, the distribution
in  Table-\ref{tab:bimar} will  be  compatible with  $\mathcal{GM}(c)$
only when at least one of  the above four conditions is violated. Interestingly, this
condition along with the E-principle is sufficient to rule out the physical realizability of $\mathcal{GM}(c)$ as discussed in next section. 

\emph{Unphysicality of GM correlation} -- We  are now in a position to
show that the correlation $\mathcal{GM}(c)$  is an unphysical one. For
this  we  consider  the  following four  different  sets  of  pairwise
exclusive events:
\begin{subequations}
	\begin{align}
\mathcal{S}_1  := & \{ (11|x^A_1x^B_1), (10|x^A_2x^B_1), (00|x^A_1x^A_2) \}, \label{eq:lo1} \\
\mathcal{S}_2  := & \{ (11|x^A_2x^B_2), (10|x^A_3x^B_2), (00|x^A_2x^A_3) \}, \label{eq:lo2} \\
\mathcal{S}_3  := & \{ (11|x^A_1x^B_1), (10|x^A_3x^B_1), (00|x^A_1x^A_3) \}, \label{eq:lo3} \\
\mathcal{S}_4  := & \{ (01|x^A_1x^A_2), (01|x^A_2x^A_3), (10|x^A_1x^A_3) \}. \label{eq:lo4} 
\end{align}
\end{subequations}
Note that here we consider events  with two inputs from same side (each
of the last events  in the first three sets and all the events in set
$\mathcal{S}_4$). This is  not illegitimate in the GM  scenario as the
bi-joint  distributions on  each side  do exist.  However, considering
such  kind  of   events  in  the  CHSH   scenario  while  establishing
unphysicality   of  post quantum  correlation   (like  PR   box)  is
illegitimate  as bi-joint  distribution  on one  side  does not  exist
in that scenario.

E-principle while  applies to first  three sets of events  implies the
constraints  $\alpha,\beta,\gamma  \leq  (1-3c)/6$ that  all  together
further imply
\begin{equation}
\alpha+\beta+\gamma  \leq  (1-3c)/2.\label{up:1}
\end{equation}
The same E-principle while applies to the set $\mathcal{S}_4$, implies,
\begin{equation}
\alpha+\beta+\gamma  \geq  (1-3c)/2.\label{up:2}
\end{equation}
Physicality      of     GM      correlation     therefore      demands
$\alpha+\beta+\gamma=(1-3c)/2$. But  it contradicts the  property (ii)
of $\mathcal{GM}(c)$ which  tells that at least one  of the conditions
in  Eq.(\ref{eq-3})  must  be   violated  for  GM  correlation.  Hence
according to the E-principle the GM  correlation cannot be a physical
one. It  is worth mentioning  that in  $1996$ Horodecki and Horodecki proved
that no quantum state of  two spin-$1/2$ particles subjected to direct
spin  measurements  generates  $\mathcal{GM}(c)$. But,  they  did  not
exclude the  possibility of physical  meaning of the  GM distributions
with  more sophisticated  experiments.  But, our  work constitute  the
strongest possible  no-go result in  this direction as it  negates any
such possibility.

\emph{Discussions}  --  Specker  first  pointed  out  that  there  are
probability theories  consistent with Kolmogorov’s axioms  that do not
satisfy the E-principle  \cite{Spek60}. Later Cabello has brought  it out in
notice  that   he  actually  considered  E-principle   as  a  fundamental principle of quantum theory \cite{Cab14}. While implications of
this  principle   in  explaining   limited  behavior   of  nonlocality
(bipartite  and multipartite  scenario) and  contextuality in  quantum
theory, here  we studied  it in the  simplest scenario  involving only
three inputs each  with two outcomes. In this scenario,  we have found
all the  restrictions imposed by  this principle at single  copy level
which need to be satisfied by a general ND probability distribution to
be physical.  However, satisfaction  of all  these restrictions
does not guarantees physicality of the  probability box. Moreover, as shown in Ref.\cite{Acin2015}, even satisfaction of the hierarchy  of  restrictions imposed  by  this principle  at
multi-copy   level does not guarantees physicality of a probability box.  An interesting  research  direction for  future may  be
finding  possible connection  of this research  to  quantum logic,  in
particular to the  recent result of Ref.\cite{Fri16}. As  an application of
our study we  show the unphysicality of GM correlation.  The method we
apply to do so also establishes a kind of limitation to the claim that
local orthogonality  principle at single  copy level is  equivalent to
no-signaling  condition \cite{FSA+13,Winter10}.  At this point it  is worthwhile  to
mention that though our result  negates physicality of GM correlation,
it does not invalidate Garg-Mermin criticism to Fine's suggestions. GM
actually  addressed  a  different  question: given  any  set  of  pair
distributions for  various pairs  of two-valued random  variables, can
there exist  a single higher  order distribution that returns  all the
pairs as marginals?  Whereas Fine result established  that the Bell-CH
inequalities  are necessary  and sufficient  for such  a higher  order
distribution when there  are just four pair  distributions (two inputs
at  each  side),  GM  construction  proved  insufficiency  of  Bell-CH
inequalities when there  are nine pair distributions  (three inputs at
each side). The unphysicality of
Popescu-Rohrlich \cite{PR94}  correlation  has  initiated  an  interesting  research
direction that motivated people to introduce    various   new
principles \cite{popescu2014nonlocality}. Unphysicality of nine pair GM distribution as established in this work may also lead to some new inquisitive results. It will be also instructive to find whether Garg-Mermin criticism of Fine's suggestion can be established by a correlation realizable in quantum world.  

{\bf Acknowledgments:} The authors would  like to thank Guruprasad Kar, Sibasish Ghosh 
and R. Srikanth for many  stimulating discussions. We would like to gratefully acknowledge Tobias Fritz for pointing out their relevant study 
(reference \cite{Acin2015}) that $\mathcal{E}^k$'s are not convex sets in general. 
We would also like to thank Arthur Fine for bringing the relevant study of reference \cite{AFine} in our notice. Private communication with N. D. Mermin
is also greatefully acknowledged. AM would like to
acknowledge  the  visit at  The  Institute  of Mathematical  Sciences,
Chennai, where  the work has  been done. AM acknowledges  support from
the CSIR project 09/093(0148)/2012-EMR-I.
    
    
%

\appendix
\section{Appendix: No-Disturbance polytope -- $\mathcal{ND}$ }

Consider a monopartite system with three inputs $\{x_1,x_2,x_3\}$ with two 
outputs $o\in\{0,1\}$ for each observable, with bi-distribution of the kind 
$\{P(o_i,o_j | x_i x_j)\}_{i<j}$, such that $ i,j\in\{1,2,3\} $ and $\{ o_i,o_j\} \in \{0,1\}$, exists. 
The normalization condition, 
\begin{equation}
\sum_{o_i,o_j} P(o_i,o_j | x_i x_j) = 1, \forall x_i,x_j
\end{equation}
and no-disturbance (ND) condition, 
\begin{equation}
P(o_1|x_1):=\sum_{o_2}P(o_1o_2|x_1x_2)=\sum_{o_3}P(o_1o_3|x_1x_3), \forall x_i,x_j,
\end{equation} 
together contributes six independent constraints with three constraints from each. 
Thus the set of these probabilities forms a 6 dimensional  convex polytope $\mathcal{ND}$ with $12$ vertices. Out of $12$ vertices of $\mathcal{ND}$, $8$ are
deterministic boxes $\{\mathcal{D}_i~|~i=1,...,8\}$ which assigns definite values to each input, and $4$ are 
indeterministic boxes $\{\mathcal{I}_j~|~j=1,...,4\}$. These deterministic 
boxes are given in Table \ref{tab:det}, and indeterministic ones are given in Table \ref{tab:indet}.

The Exclusivity principle at single copy level imposes the following sets of non-trivial constraints:
\begin{subequations}
	\begin{align}
P(00|x_1x_2)+P(10|x_2x_3)+P(11|x_1x_3)\le1,\label{eq:ap1}\\
P(00|x_1x_2)+P(11|x_2x_3)+P(10|x_1x_3)\le1,\label{eq:ap2}\\
P(01|x_1x_2)+P(00|x_2x_3)+P(11|x_1x_3)\le1,\label{eq:ap3}\\
P(01|x_1x_2)+P(01|x_2x_3)+P(10|x_1x_3)\le1,\label{eq:ap4}\\
P(10|x_1x_2)+P(10|x_2x_3)+P(01|x_1x_3)\le1,\label{eq:ap5}\\
P(10|x_1x_2)+P(11|x_2x_3)+P(00|x_1x_3)\le1,\label{eq:ap6}\\
P(11|x_1x_2)+P(00|x_2x_3)+P(01|x_1x_3)\le1,\label{eq:ap7}\\
P(11|x_1x_2)+P(01|x_2x_3)+P(00|x_1x_3)\le1\label{eq:ap8}.
\end{align}
\end{subequations}
All of the eight deterministic extremal points of $\mathcal{ND}$ satisfy all of the constraints, i.e., Eqs.(\ref{eq:ap1})-(\ref{eq:ap8}). On the other hand, each of the indeterministic boxes violates two of theses above constraints. For example the indeterministic box $\mathcal{I}_1$ violates the constraints (\ref{eq:ap2}) and (\ref{eq:ap7}). Note that the violation amount goes up to $3/2$, and this is the maximal violation of the above constraints by any ND probability box. If we consider mixture of indeterministic box $\mathcal{I}_1$ and completely random box $\mathcal{W}$, i.e., $p\mathcal{I}_1+(1-p)\mathcal{W}$, then the constraints (\ref{eq:ap2}) and (\ref{eq:ap7}) will be violated whenever $p>1/3$.  

\begin{table}[h!]
	\begin{tabular}{c|| c| c| c| c| c| c| c| c|}
		& $\mathcal{D}_1$ & $\mathcal{D}_2$ & $\mathcal{D}_3$ &$\mathcal{D}_4$ & $\mathcal{D}_5$ & $\mathcal{D}_6$ & $\mathcal{D}_7$ & $\mathcal{D}_8$ \\
		\hline
		$P_{00|x_1x_2}$ & 1 & 1 & 0 & 0 & 0 & 0 & 0 & 0 \\
		$P_{01|x_1x_2}$ & 0 & 0 & 1 & 0 & 1 & 0 & 0 & 0 \\
		$P_{10|x_1x_2}$ & 0 & 0 & 0 & 1 & 0 & 1 & 0 & 0 \\
		$P_{11|x_1x_2}$ & 0 & 0 & 0 & 0 & 0 & 0 & 1 & 1 \\
		\hline
		$P_{00|x_2x_3}$ & 1 & 0 & 0 & 1 & 0 & 0 & 0 & 0 \\
		$P_{01|x_2x_3}$ & 0 & 1 & 0 & 0 & 0 & 1 & 0 & 0 \\
		$P_{10|x_2x_3}$ & 0 & 0 & 1 & 0 & 0 & 0 & 1 & 0 \\
		$P_{11|x_2x_3}$ & 0 & 0 & 0 & 0 & 1 & 0 & 0 & 1 \\
		\hline
		$P_{00|x_1x_3}$ & 1 & 0 & 1 & 0 & 0 & 0 & 0 & 0 \\
		$P_{01|x_1x_3}$ & 0 & 1 & 0 & 0 & 1 & 0 & 0 & 0 \\
		$P_{10|x_1x_3}$ & 0 & 0 & 0 & 1 & 0 & 0 & 1 & 0 \\
		$P_{11|x_1x_3}$ & 0 & 0 & 0 & 0 & 0 & 1 & 0 & 1 \\
		\hline
	\end{tabular}
	\caption{The $8$ deterministic boxes $\mathcal{D}_i$. Tri-joint probability distributions exists for each of the boxes.}
	\label{tab:det}
\end{table}

\begin{table}[h!]
	\begin{tabular}{c|| c| c| c| c|}
		& $\mathcal{I}_1$ & $\mathcal{I}_2$ & $\mathcal{I}_3$ & $\mathcal{I}_4$\\
		\hline
		$P_{00|x_1x_2}$ & $1/2$ & $1/2$ & 0 & 0 \\
		$P_{01|x_1x_2}$ & 0 & 0 & $1/2$ & $1/2$  \\
		$P_{10|x_1x_2}$ & 0 & 0 & $1/2$ & $1/2$ \\
		$P_{11|x_1x_2}$ & $1/2$ & $1/2$ & 0 & 0  \\
		\hline
		$P_{00|x_2x_3}$ & $1/2$ & 0 & $1/2$ & 0 \\
		$P_{01|x_2x_3}$ & 0 & $1/2$ & 0 & $1/2$  \\
		$P_{10|x_2x_3}$ & 0 & $1/2$ & 0 & $1/2$  \\
		$P_{11|x_2x_3}$ & $1/2$ & 0 & $1/2$ & 0  \\
		\hline
		$P_{00|x_1x_3}$ & 0 &$1/2$ & $1/2$ & 0  \\
		$P_{01|x_1x_3}$ & $1/2$ & 0 & 0 & $1/2$  \\
		$P_{10|x_1x_3}$ & $1/2$ & 0 & 0 & $1/2$  \\
		$P_{11|x_1x_3}$ & 0 & $1/2$ & $1/2$ & 0  \\
		\hline
	\end{tabular}
	\caption{The $4$ indeterministic boxes $\mathcal{I}_j$. No tri-joint probability distributions exists for any of theses boxes.}
	\label{tab:indet}
\end{table}

\end{document}